\documentclass[conference]{IEEEtran}

\usepackage{verbatim} 

\usepackage{color} 
\usepackage{url} 

\usepackage[caption=false,font=footnotesize]{subfig}
\usepackage{fixltx2e}
\usepackage[cmex10]{amsmath}
\newtheorem{definition}{Definition}

\usepackage[dvips]{graphicx}
\graphicspath{{./eps/}}
\DeclareGraphicsExtensions{.eps}

\usepackage{cite}

\begin{document}

\title{Detecting Flow Anomalies in Distributed Systems}

\author{
	\IEEEauthorblockN{Freddy C. Chua}
	\IEEEauthorblockA{Mechanisms and Design Lab\\
	HPLabs, Palo Alto, CA 94304, USA\\
	freddy.chua@hp.com}
	\and
	\IEEEauthorblockN{Ee-Peng Lim}
	\IEEEauthorblockA{School of Information Systems\\
	Singapore Management University\\
	eplim@smu.edu.sg}
	\and
	\IEEEauthorblockN{Bernardo A. Huberman}
	\IEEEauthorblockA{Mechanisms and Design Lab\\
	HPLabs, Palo Alto, CA 94304, USA\\
	bernardo.huberman@hp.com}
}

\maketitle

\begin{abstract}
Deep within the networks of distributed systems, one often finds anomalies that affect their efficiency and performance. These anomalies are difficult to detect because distributed systems may not have sufficient sensors to monitor the traffic flow within their interconnected nodes. Without early detection and corrections, these anomalies can aggravate over time and possibly cause disastrous outcomes in the system in an unforeseeable future. Using only coarse-grained information from the two end points of network flows, we developed a network transmission model and localization algorithm that detects and ranks the location of anomalies. We evaluate our approach using passengers' records of an urbanized city's public transportation system, and correlate our findings with passengers' postings on social media microblogs. Our experiments show that our localization algorithm gives a better ranking of anomalies than standard deviation measures drawn from statistical models. Our case study also demonstrates that transportation events reported in social media microblogs often match the locations of our detected anomalies detected with our algorithm.
\end{abstract}

\section{Introduction}
\label{sec:intro}

In a complex world, networks offer a useful abstract representation for organizing the relationships between entities of interest in distributed systems. Entities are represented as nodes while edges connecting pairs of nodes represent relationships between them. Examples of pervasive distributed systems are social networks \cite{Wu2004}, protein networks, computer networks \cite{Kind2009, Sengar2009}, transportation networks \cite{Fadlil2013, Yuan2013}, logistical networks \cite{Agovic2009}, neurological networks, organizational networks \cite{Mihm2010}, wireless sensor networks (Internet of Things), electrical networks, and many more.

A functional network requires reliable and consistent flow of entities through its if it is to achieve its objectives. However, it is inevitable that the building blocks of the system deteriorate non-uniformly over time, leading to occasional anomalous behavior in certain parts of the system. Anomalies in such systems can disrupt normal operations and prevent the network from meeting its objectives in a timely manner.

While critical anomalies leading to catastrophic failures are noticed and addressed by the stakeholders of the distributed system, it is more challenging to recognize the \emph{non-critical} ones that result in a lower than optimal efficiency of the system. Since in the latter case the system can continue to function without corrections, non-critical anomalies are often hard to locate and \emph{ignored}. But if not corrected, non-critical anomalies can aggravate over time and lead to the catastrophic failures of the system in an unforeseeable future. 

\begin{figure}[htb]
	\centering
	\includegraphics[width=3.5in]{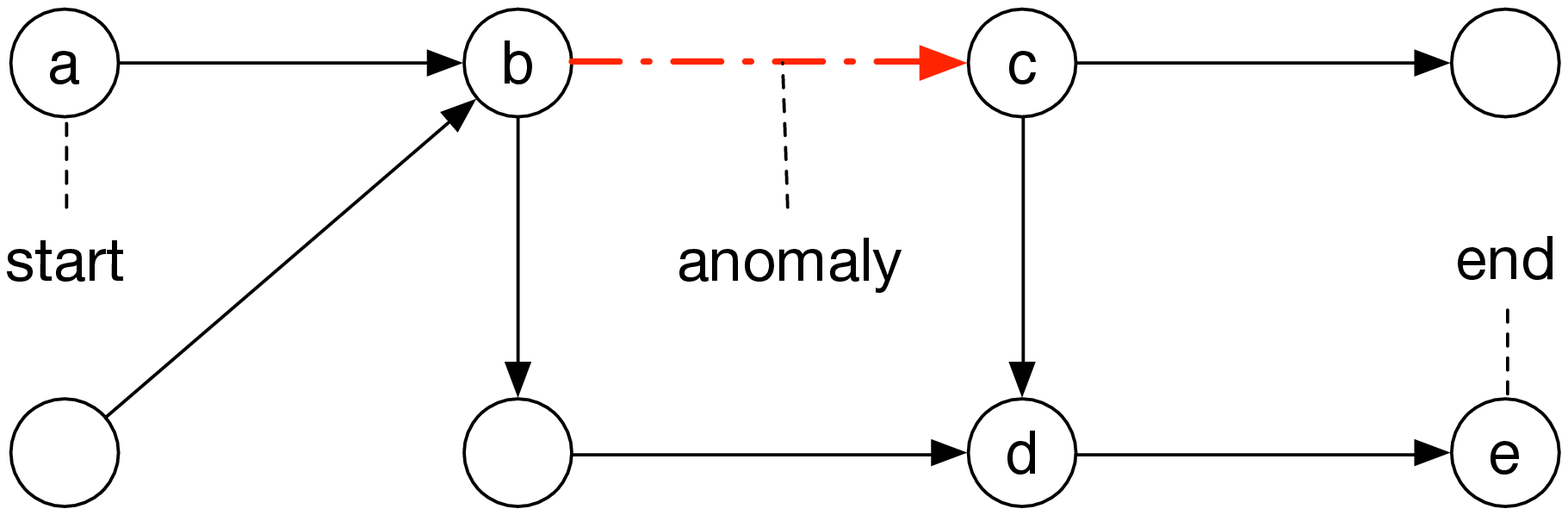}
	\caption{Entities Flow in Networks}
	\label{fig:travel_graph}
\end{figure}

\begin{figure}[htb]
	\centering
	\subfloat[Histogram at the start node of an entity-flow route]{
		\includegraphics[width=3.5in]{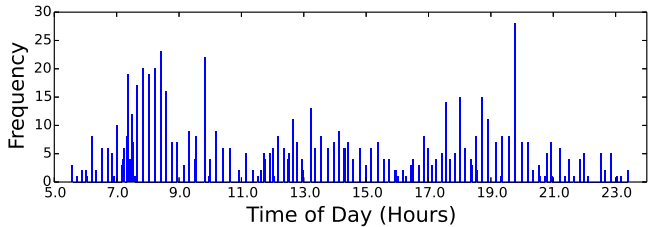}
		\label{fig:7_start}
	}\\
	\subfloat[Histogram at the end node of an entity-flow route]{
		\includegraphics[width=3.5in]{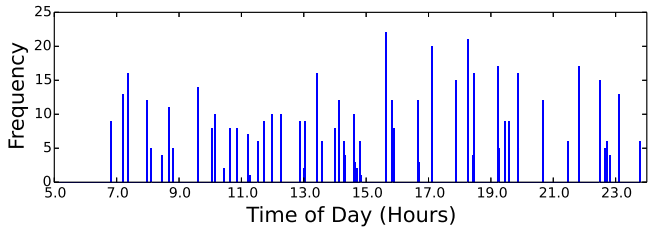}
		\label{fig:7_end}
	}
	\caption{Histograms for start and end nodes of a flow route}
	\label{fig:7}
\end{figure}

Before proceeding further, we use Figures \ref{fig:travel_graph} and \ref{fig:7} to illustrate the problem. Figure \ref{fig:travel_graph} shows a  distributed system where entities flow from node to node through directed edges. The edges connecting nodes $\{a, b, c, d, e\}$ form a route through which the entities flow. We do not assume that $a$ and $e$ are always the origin and destination of every entity flow, i.e. entities within the distributed system could originate from or terminate at any of the intermediate nodes $\{b, c, d\}$. The \textcolor{red}{dotted $- ~ \cdot \rightarrow$} line indicates the possibility of an existing anomaly that would disrupt the regular flow of entities along this route.

Figures \ref{fig:7_start} and \ref{fig:7_end} show the histograms of entities starting its flow at the origin node $a$, and ending its flow at the destination node $e$. In the histograms, the x-axis represents the hour of the day, and each bin on the x-axis has a time interval of two minutes. The y-axis shows the number of flows starting or ending at the time corresponding to the bins. The phenomenon that we could immediately observe is that while the start node shows a \emph{regular} transmission of entities, the end node receives the entities at \emph{irregular} intervals. This could suggest the presence of an anomaly within the path such as the segment connecting node $b$ and $c$, suffering from a severe network congestion as shown in the example of Figure \ref{fig:travel_graph}.

In this paper we propose a \emph{non-intrusive solution} to early detection of such anomalies. Our proposed non- solution relies on \emph{temporal} data related to the \emph{flow} of entities from an \emph{origin} node to a \emph{destination} node. Since the solution should be non-intrusive, we only require temporal information from the two (origin and destination) end points while assuming that detailed knowledge of the flow through the intermediate nodes of its path is missing or difficult to obtain.

We formally define our problem based on the assumption that the following recorded data is available for our analysis. That is, given a set of records $R$ of a distributed system, each record $r \in R$ contains the following,
\begin{enumerate}
	\item Spatial: The origin node $x_r$, and destination node $y_r$ of entity flow in $r$.
	\item Temporal: The time $t(x_r)$ when entity flow starts at the origin and the time $t(y_r)$ when entity flow ends at the destination.
	\item Cost: The distance $d_r$ from $x_r$ to $y_r$ traveled by the entity, or the non-temporal cost incurred due to the entity flow.
	\item The path $p_r$ taken by $r$. The path consists of the sequence of nodes that the entity visits for it to flow from node $x_r$ to node $y_r$. In situations where complete knowledge of the network or path is not known, it would still be possible to infer the path based on the distance traveled.
\end{enumerate}
A pair of consecutive nodes $i,j$ in path $p_r$, forms a segment $s_{i,j}$. We determine whether the observed amount of time $\hat{t}_r$, taken for entity flow in $p_r$ deviates significantly from the expected amount of time $t_r$. For all the records $r \in R$ with observed time that deviates significantly from the expected time, we locate the segments $s_{i,j} \in p_r$ that are likely to be the cause of the deviations. 

This task is challenging due to the lack of knowledge on the time it takes for entities to flow through the individual segments of path $p_r$. We need to infer the expected time for each segment based on the set of available records we have and the limited amount of knowledge each record contains.

Our technical contribution consists of the proposal of a network transmission model that performs the inference on spatial-temporal data where knowledge of the temporal trajectory is missing or not available. Such situations are common when it is not possible to install sensors for monitoring the internal networks of distributed systems. For example, in computer networks, the application layer only have knowledge of the two endpoints, and does not know the behavior of the intermediate nodes. Another example is transportation networks, which only records the passengers boarding and alighting stations for the purpose of calculating their transportation fare. 

Building on the network transmission model, we further propose an algorithm that ranks the anomalous data in order of importance by measuring how much impact each anomalous data has on others. Then using the ranking, we can isolate the location of where the anomalies occur in the distributed system, listed in descending order of importance. We test our models and algorithms on the network data of a physical transportation system and verify the accuracy of the detected anomalies through case studies using social media data.

We present an overview of prior work on anomalies detection in Section \ref{sec:related}. Section \ref{sec:models} describes the details of our proposed network transmission models as well as two baselines for comparison. Using the results of our proposed model, we utilize the algorithm as described in Section \ref{sec:localization} for locating the segments of where the anomalies might have occurred. Then we evaluate the performance of our proposed models and the localization algorithm in Section \ref{sec:experiments}. We conclude our work in Section \ref{sec:conclusion} and highlight certain possible extensions that can be made for this area of research.

\section{Related Work}
\label{sec:related}

\textbf{Overview of Anomaly Detection.} Chandola et al. \cite{Chandola2009} gave a comprehensive survey on the topic of anomaly detection for general scenarios. Chandola et al. defined an anomaly as a pattern that does not conform to expected normal behavior. But the notion of expected normal behavior depends on the application domains and types of input data. Chandola et al. surveyed a broad overview of various techniques used in anomaly detection; classification-based \cite{Stefano2000, Barbara2001}, nearest neighbor approach \cite{Otey2006}, clustering-based \cite{He2003}, statistical-based (including parametric \cite{Eskin2000} and non-parametric \cite{Chow2002}), information-theoretic-based \cite{Ando2007}, and spectral anomaly detection techniques \cite{Agovic2009}. Although our network transmission model can be classified as one of these methods, the second portion of our paper that measures the impact of anomalous data has not been used before. The kind of input data we used for our anomalies detection is also unique and contains inherent difficulties, which is addressed by our network transmission model.

\textbf{Job Anomaly Detection.} Fu et al. \cite{Fu2009} addressed the detection for two broad classes of anomalies; The first class is work flow execution anomalies and the second class is execution low performance anomalies. Our focus on non-critical anomalies is similar to Fu et al.'s definition of execution low performance anomalies. Their detection method is based on text analysis of logs generated by parallel frameworks such as Hadoop. However, we do not require the usage of logs, which is usually difficult to obtain from distributed systems.

\textbf{Computer Network Anomaly Detection.} \cite{Kind2009, Sengar2009} proposed anomalies detection methods specifically for computer networks. Such anomalies could come from hackers that infiltrate the network and attempt to compromise the security of the computer network. Kind et al. \cite{Kind2009} suggested that, the existence of anomalies may not necessarily cause significant changes in the performance and speed of network flow. They proposed a feature-based anomaly detection method that uses the information in headers of computer network packets. Sengar et al. \cite{Sengar2009} proposed a behavioral distance metric that is adaptable for online detection of streaming network packets. While the kind of anomalies Kind et al. \cite{Kind2009} and Sengar et al. \cite{Sengar2009} studied are also elusive in nature, it only applies to a subset of distributed systems such as computer networks with security vulnerabilities. On the other hand, the models and algorithms we propose in our paper is well suited for many kinds of distributed systems that do not have the same kind of security flaws found in computer networks.

\textbf{Anomaly Detection in Transportation.} Agovic et al. \cite{Agovic2009} proposed a manifold embedding based method for detecting anomalies in transportation. Their algorithm takes in high dimensional feature vectors and reduces it to low dimensional representations for better efficiency of detecting anomalies. However, in the network data we look at in this research, our data is low dimensional and coarse-grained information of network flow in distributed systems. We use low dimensional information and reconstruct the high dimensional information in order to obtain a better representation of the network flows.

Fadlil et al. \cite{Fadlil2013} used data that has multiple sensors monitoring different variables of the road conditions. The multiple sensors provided a multi-view of readings that is high dimensional which required manifold embedding \cite{Agovic2009} for clustering of the data points into two clusters. From the two clusters, Fadlil et al. chose the smaller cluster as the set of anomalous data points. But their approach requires detailed data such as readings from multiple sensors. On the other hand, our algorithm is suitable for distributed systems that do not have access to many sensors.

Thottan and Ji \cite{Thottan2003} defined network anomalies as network operations that deviate from normal network behavior. They proposed a method of network anomalies detection by monitoring several variables of network operations over time, and model the time series as an autoregressive process. The presence of abrupt changes in the time series is detected as anomalies. In our case, apart from detecting statistical deviations, we go one step further to detect the location of where the anomalies occur in the distributed system.

Pan et al. \cite{Pan2013} address the problem of detecting and describing traffic anomalies using crowd sensing with Beijing's taxis' GPS data and China's Weibo\footnote{This service resembles Twitter but is catered for the Chinese population in Chinese language.} social media data. The anomaly here refers to a deviation in traffic volume on segments of road during some special events. The proposed detection algorithm is straightforward because of the availability of GPS data at regular and fine grain time intervals. Their focus is on two special indexing data structures that improve the algorithm efficiency.

Liu et al. \cite{Liu2011} proposed the discovery of causal interactions among traffic outliers. The proposed framework first partitions the geographical space into regions represented by nodes in a region graph. Edges between nodes represent the traffic flow between regions. By comparing the features of each edge across different time frames, Liu et al. is able to identify the outlying travel trajectories. From the detected outliers, an outlier tree is constructed with each node representing an outlier trajectory. The parent of an outlier node occurs before in time and the destination of the parent is said to be the cause or origin of the child.

\cite{Ge2011,Zhang2011,Zhang2012} analyzed taxi GPS data to detect drivers who overcharge their passengers by deliberately taking the longer route to reach the destination. The general idea for finding these anomalous routes is to compare the route taken for each pickup and destination points and obtain a measure of how much it deviates from the usual routes.

Chawla et al. \cite{Chawla2012} proposed an algorithm to detect anomaly based on Pinciple Component Analysis (PCA). Chawla et al. \cite{Chawla2012} represent the traffic data into two matrices, 1) the link-path matrix and the 2) link-time matrix. Then using PCA, Chawla et al. \cite{Chawla2012} is able to factorize the matrices into eigenvectors with its respective eigenvalues. The eigenvectors corresponding to large eigenvalues represent the norm, while those eigenvectors corresponding to lower eigenvalues represent the anomalies. Using these anomalies, Chawla et al. \cite{Chawla2012} tested their method to see if they could determine the root cause on synthetically generated data sets.

All of these previous works \cite{Chawla2012, Ge2011, Liu2011, Pan2013, Zhang2011, Zhang2012} require fine-grained sampling of GPS and time for anomaly detection and is limited only to spatial-temporal situations such as transportation networks. Our approach does not require GPS data and is made general for most forms of network traffic in distributed systems.

\section{Network Transmission Models}
\label{sec:models}

To find anomalies in the network, we analyze the set of entities flow records $R$ from a distributed system. But we must first determine which recorded entity flow $r \in R$ is anomalous before we could proceed further with the localization task. A record $r$ is anomalous, if the observed time $\hat{t}_r$ taken to complete the distance $d_r$ deviates significantly from the expected value $t_r$ given to us by a statistical model.

\subsection{Baseline 1}

For our first baseline, we assume that every recorded entity flow $r$ travels at a constant speed $c$ to cover the total distance $d_r$ required to reach its destination. The distribution of time taken $t_r$ for record $r$ is given by the following Gaussian distribution,
\[ t_r \sim \mathcal{N} \left( \frac{d_r}{c}, d_r \sigma^2 \right) \]
The unknown value of $c$ can be obtained by minimizing the following Sum-of-Squares error,
\[ \sum_{r \in R} (t_r - \hat{t}_r)^2 \]
Thus allowing us to obtain,
\begin{gather*}
	c = \frac{\sum_{r \in R} d_r}{\sum_{r \in R} \hat{t}_r} \qquad
    \sigma^2 = \frac{ \sum_{r \in R} \left( \hat{t}_r - \frac{d_r}{c} \right)^2 }{ \sum_{r \in R} d_r }
\end{gather*}

\subsection{Baseline 2}

The second baseline model which provides a more discriminative estimation than the first baseline is to assume a speed $c_p$ for each distinct path $p$ in the network.
\[ t_r \sim \mathcal{N} \left( \frac{d_r}{c_p}, d_r \sigma^2 \right) \]
The estimation for $c_p$ and $\sigma^2$ can be easily extended from the first baseline to obtain the following,
\begin{gather*}
	c_p = \frac{\sum_{r \in R_p} d_r}{\sum_{r \in R_p} \hat{t}_r} \qquad
    \sigma^2 = \frac{ \sum_{p \in P} \sum_{r \in R_p} \left( \hat{t}_r - \frac{d_r}{c_p} \right)^2 }{ \sum_{r \in R} d_r }
\end{gather*}
where $P$ is the set of possible paths and $R_p$ is the set of records that took the path $p \in P$.

\subsection{Edge-based Model}

We propose the Edge-based model that models the speed of individual edges in the network. Given an entity flow record $r$ that originates from $x_r$ and terminates at $y_r$, the path taken by $r$ is denoted as $p_r$. Within each path $p_r$, the entity flows through consecutive sequences of nodes. For every pair of consecutive nodes $i,j \in p_r$, the segment $s_{i,j}$ connecting the pair is associated with the known distance $d_{i,j}$ and an estimated speed $c_{i,j}$. The time taken $t_{i,j}$ for each segment $s_{i,j}$ could then be estimated using the distance $d_{i,j}$ and speed $c_{i,j}$. Summation of the estimated time in each segment gives the expected time $t_r$ needed for $r$ to travel from $x_r$ to $y_r$.

The distribution of time for each segment $s_{i,j}$ is given by the following Gaussian distribution,
\[ t_{i,j} \sim \mathcal{N} \left( \frac{d_{i,j}}{c_{i,j}}, d_{i,j} \sigma^2 \right) \]

The distribution of time $t_r$ of $r$ is given by the following linear Gaussian distribution,
\begin{align*}
	t_r &= \sum_{ (i,j) \in p_r } t_{i,j} \\
	t_r &\sim \mathcal{N} \left( \sum_{ (i,j) \in p_r } \frac{ d_{i,j} }{ c_{i,j} }, d_r \sigma^2 \right)
\end{align*}
To estimate the variance $\sigma^2$,
\begin{align*}
	\sigma^2 = \frac{ \sum_{r \in R} \left( \hat{t}_r - \sum_{ (i,j) \in p_r } \frac{d_{i,j}}{c_{i,j}} \right)^2}{\sum_{r \in R} d_r}
\end{align*}
To estimate the speed $c_{i,j}$ of every segment $s_{i,j}$ using the observed time $\hat{t}_r$ of each record $r$, we \textbf{maximize} the log likelihood $\mathcal{L}_r$ from each $r$. The log likelihood $\mathcal{L}_r$ as contributed by $r$ is given by,
\begin{align*}
	\mathcal{L}_r &= \log \left( \frac{1}{\sqrt{2 \pi d_r \sigma^2 }} \right) - \frac{\left( \hat{t}_r - \sum_{ (i,j) \in p_r } \frac{d_{i,j}}{c_{i,j}} \right)^2}{2 d_r \sigma^2 } \\
	&= - \frac{1}{2} \log \left( d_r \sigma^2 \right) - \frac{\left( \hat{t}_r - \sum_{ (i,j) \in p_r } \frac{d_{i,j}}{c_{i,j}} \right)^2}{2 d_r \sigma^2 }
\end{align*}
We add a log barrier penalty to prevent negative speeds. A larger $c_{i,j}$ increases the $\mathcal{L}^*_r$ term as given by the addition, which is encouraged since we are trying to maximize the log likelihood.
\begin{align}
	\label{eqn:objective1}
	\mathcal{L}^*_r = \mathcal{L}_r + \tau \sum_{(i,j) \in p_r} \log c_{i,j}
\end{align}
where $\tau$ is the strength of the penalty. By taking partial derivative with respect to $c_{p,q}$,
\begin{align}
	\label{eqn:sgd_mean}
	\frac{\partial \mathcal{L}^*_r}{\partial c_{p,q}} = - \frac{\left( \hat{t}_r - \sum_{ (i,j) \in p_r } \frac{d_{i,j}}{c_{i,j}} \right)}{d_r \sigma^2 } \cdot \frac{d_{p,q}}{c_{p,q}^2} + \frac{\tau}{c_{p,q}}
\end{align}
The partial derivative allows us to perform Stochastic Gradient Descent (SGD) on parameters $c_{i,j}$ as follows,
\begin{align*}
	c_{i,j} \leftarrow c_{i,j} + \eta \frac{\partial \mathcal{L}^*_r}{\partial c_{i,j}}
\end{align*}
There are several interesting properties with the partial derivative in Equation \ref{eqn:sgd_mean}. The variance in the denominator shows that the more uncertain we are, the lesser the gradient is, hence less changes to $c_{i,j}$. The smaller $c_{i,j}$ is, the second component in Equation \ref{eqn:sgd_mean} will compensate by adding positive value to prevent $c_{i,j}$ from entering the negative region.

\subsection{Smoothed Edge-based Model}

To avoid overfitting the model parameters to the observed data set, we add additional constraints to Equation \ref{eqn:objective1} that minimize the difference between the speeds of consecutive segments in a path. This constraint is based on the assumption that consecutive segments have related speeds. In the equation that follows, a larger difference between the speeds of two consecutive segments lowers the log likelihood as given by the subtraction, which is discouraged.
\begin{align*}
	\mathcal{L}^{**}_r = \mathcal{L}^*_r - \frac{\psi}{2} \sum_{(i,j) \in p_r} \left( c_{i,j} - c_{j,k} \right)^2
\end{align*}
where $c_{j,k}$ is speed of $s_{j,k}$ that comes after $s_{i,j}$.

Estimation for the variance $\sigma^2$ remains the same while estimation of $c_{i,j}$ is slightly modified,
\begin{align*}
	\frac{\partial \mathcal{L}^{**}_r}{\partial c_{i,j}} &= \frac{\partial \mathcal{L}^*_r}{\partial c_{i,j}} - \psi (c_{i,j} - c_{j,k}) \\
	c_{i,j} &\leftarrow c_{i,j} + \eta \frac{\partial \mathcal{L}^{**}_r}{\partial c_{i,j}}
\end{align*}

In Section \ref{sec:experiments}, we would evaluate which of these models is a better choice in terms of fitting to records that are not observed during the estimation (training) phase.

\section{Localization of Network Anomalies}
\label{sec:localization}

The models as described in the previous section would allow us to determine whether a record $r \in R$ is anomalous by comparing the difference of the observed time and the expected time $\hat{t}_r - t_r$ with the standard deviation $\sqrt{d_r \sigma^2}$. We use the following ratio $\alpha_r$ to measure the degree of deviation.
\begin{align}
	\label{eqn:alpha_ratio}
	\alpha_r = \frac{\hat{t}_r - t_r}{\sigma \sqrt{d_r}}
\end{align}

Given any record $r \in R$, $\alpha_r > 1$ indicates that the time taken is longer than expected while $\alpha_r < 1$ indicates that the time taken is shorter than expected. In most distributed systems, the records of interest for further investigation would be those with $\alpha_r > \delta$, where $\delta$ is a cut-off value to determine whether $r$ has a significantly larger observed time than expected. We would be able to obtain a reduced set of records $R_{\alpha > \delta}$ such that $r \in R_{\alpha > \delta}$ has a ratio $\alpha_r > \delta$. Using the reduced set $R_{\alpha > \delta}$ instead of the full set $R$, we could save computational costs by focusing on a smaller set of records for finding the location of anomalies in the distributed system.

\begin{figure}
	\centering
	\includegraphics[width=3.5in]{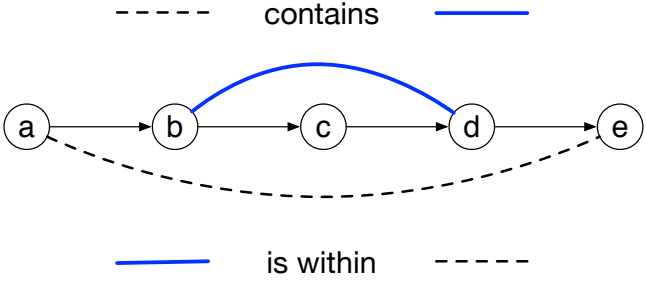}
	\caption{Example for the \emph{contains} and \emph{within} definitions}
	\label{fig:contain_within_example}
\end{figure}

But a high ratio $\alpha_r$ for record $r \in R_{\alpha > \delta}$ could be an isolated incident that does not have any significant impact on the distributed system. The ratio $\alpha_r$ also does not reveal the specific segment $s_{i,j}$ in the path $p_r$ of $r$ that causes the longer observed time $\hat{t}_r$. To address these issues, we propose an algorithm that serves two purposes:
\begin{enumerate}
	\item Measuring how many other records $r' \in R_{\alpha > \delta}\setminus{r}$ are related to $r$ in order to determine the significance of the network congestion in the path $p_r$ taken by $r$.
	\item Locating the segment $s_{i,j} \in p_r$ that is most likely to contribute to the high $\alpha_r$ ratio of $r$.
\end{enumerate}
We first define the ``relatedness'' of two records $r$ and $r'$ in more precise terms using ``contains'' and ``within''.
\begin{definition}
	$r'$ \emph{contains} $r$ if all of the following conditions are satisfied: 
	\begin{enumerate}
		\item The path $p_{r'}$ connecting the origin $x_{r'}$ and destination $y_{r'}$ of $r'$, passes through all the nodes of path $p_{r}$ that connects the origin $x_{r}$ and destination $y_{r}$ of $r$. Figure \ref{fig:contain_within_example} shows an example of this condition. The path connecting node $a$ to node $e$ contains the path which connects node $b$ to the node $d$.
		\item The time $t(x_{r'})$ when $r'$ starts at origin $x_{r'}$, is earlier than the time $t(x_{r})$ when $r$ starts at origin $x_{r}$. i.e. $t(x_{r'}) < t(x_{r})$.
		\item The time $t(y_{r'})$ when $r'$ ends at destination $y_{r'}$, is later than the time $t(x_{r})$ when $r$ ends at destination $y_{r}$. i.e. $t(y_{r'}) > t(y_{r})$.
	\end{enumerate}
\end{definition}
\begin{definition}
	$r$ is \emph{within} $r'$ if and only if $r'$ contains $r$.
\end{definition}

Based on the two definitions, the algorithm for localizing the anomalies in the network proceeds as follows:
\begin{enumerate}
	\item Obtain the set of records $R_{\alpha > \delta}$ such that $\forall r \in R_{\alpha > \delta}$ has ratio $\alpha_r > \delta$. This gives us the set of records $R_{\alpha > \delta}$ with observed travel time that is significantly larger than the expected travel time.
	\item For each $r \in R_{\alpha > \delta}$, obtain the set of records $R_r$, where
	\[ R_{r} := \{ r' \in R_{\alpha > \delta} | r' ~ \text{contains} ~ r \land r' \neq r \} \]
	That is, $R_r$ is the set of records that contains $r$. The value of $|R_r|$ has a positive correlation on the importance of path $p_r$ to other records and traffic.
	\item Then by sorting the set of records $R_{\alpha > \delta}$ in descending order of $|R_r|, \forall r \in R_{\alpha > \delta}$, and examining the segments $s_{i,j}$ of path $p_r$, we would be able to locate the segments $s_{i,j} \in p_r$ with severe network congestion between the times of $t(x_r)$ and $t(y_r)$.
	\item For any given $r' \in R_{\alpha > \delta}$, we would also be able to locate the congested segments of path $p_{r'}$ by using the path $p_r$ of record $r$, where $r$ is within $r'$ and $R_r = \emptyset$.
\end{enumerate}

\section{Experiments}
\label{sec:experiments}

We first apply the models from Section \ref{sec:models} on the data set of a distributed system to test the generalization performance. The model that has the best generalization performance would have the lowest prediction error for unobserved data. We then select the best model to use for detection and localization of network anomalies in the distributed system by applying the algorithm as described in Section \ref{sec:localization}. To evaluate the reliability of the detected and localized anomalies, we examine a set of tweets from Twitter of the same period to check whether users of the distributed system expressed their frustrations by tweeting (complaining) on Twitter.

\subsection{Data Description}

We evaluate our models on a data set that contains the passengers' travel records on the Public Transportation System (PTS) of an urbanized city. The PTS consists of the railway system and the public bus system.

Each passenger carries a payment card containing a RFID chip. The RFID chip allows the companies to identify the passenger and charge the transportation fare to their account. The passengers boarding and alighting geo-locations are recorded for each travel trip in order to charge the appropriate amount based on the distance traveled. The time boarded and time alighted are also recorded, which makes it possible to calculate the time taken for the travel.

Being a densely populated city, majority of the city's population commute via the PTS, instead of driving in privately owned vehicles. The data set thus represents an almost complete usage of a real and large-scale distributed system, which would allow us to utilize it for verifying our proposed models and algorithms.

We mentioned in Section \ref{sec:intro} that we wanted to detect non-critical anomalies because it is less noticeable than critical anomalies. The transportation network of bus routes would contain many of these non-critical anomalies because traffic in congested road segments can be diverted to other roads. Each record $r$ contains the following information of a passenger's journey:
\begin{enumerate}
	\item Bus stop ID of boarding, $x_r$.
	\item Bus stop ID of alighting, $y_r$.
	\item Date and time of boarding, $t(x_r)$.
	\item Date and time of alighting, $t(y_r)$.
	\item Distance traveled between boarding and alighting, $d_r$.
	\item Bus service: The bus service is a number, which represents the unique route taken by the bus. Many different buses operate using the same service number so that different buses can pick up or alight passengers at various bus stops with regular intervals. Using the bus service number, the bus stop ID of boarding and alighting, we are able to obtain the path $p_r$ traveled by the passenger of this record $r$.
\end{enumerate}
The bus routes remain relatively static but may change due to road maintenance and repairs. To ensure that we always have the correct bus routes for each specific bus service, we perform a simple bus route inference step in the next section.

\subsubsection{Bus Route Inference}

\begin{figure}[htb]
	\centering
	\includegraphics[width=3.5in]{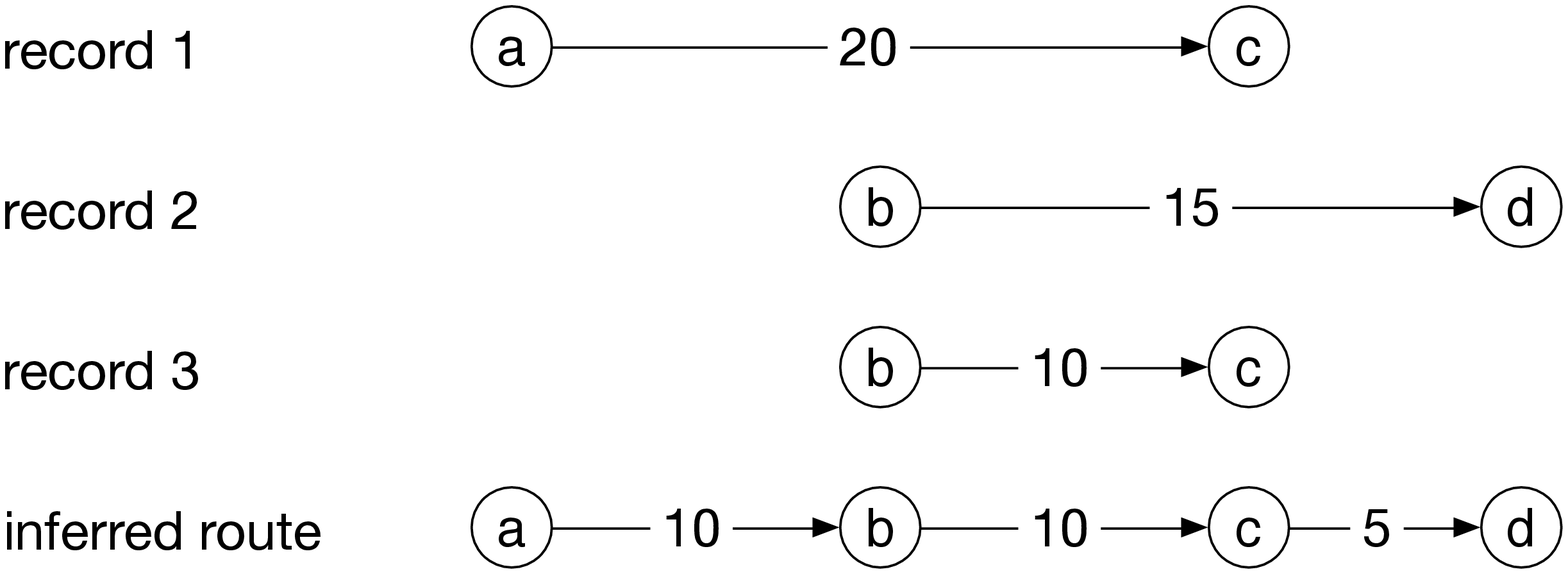}
	\caption{Examples of the trip records in the data set for a specific bus service}
	\label{fig:trip_example}
\end{figure}

Figure \ref{fig:trip_example} shows an example of three records for the \emph{same} bus service number. We have the time and location of the passenger boarding and alighting from the bus for each of their journeys. The nodes in Figure \ref{fig:trip_example} represent the bus stops while the number in the connecting edges represent the distance between two bus stops. By using different records of the same bus service, we are able to construct the route that the bus service takes, as shown in Figure \ref{fig:trip_example}. For example, from Figure \ref{fig:trip_example}, we can infer that $b$ should be between $a$ and $c$ because $b$ is nearer to $c$ compared to $a$, and $c$ is nearer to $b$ than $d$ so $d$ should come after $c$.
Then by combining the inferred routes of various bus services, we would be able to obtain a transportation network as shown in Figure \ref{fig:travel_graph}.

\subsubsection{Statistics}

We perform our initial analysis on three days of data, December 8th 2011, December 15th 2011 and December 22nd 2011. These three days are spaced one week apart and falls on Thursday of the week, which is a typical working day. The reason for choosing these specific days is because of the occurrence of an event (external to the bus system) on December 15th 2011 which affected the transportation system of the city. The choice of December 8th and December 22nd is to show that the anomalies present on December 15th, is absent one week before and one week after. We perform the bus route inference on the records of these three days and obtained the respective transportation network for each day. Statistics of the transportation network is shown in Table \ref{tbl:statistics}. While most of the transportation network remains fairly static, there are minor fluctuations in the size of the network due to several reasons: 1) The city performs road maintenance frequently, resulting in temporary changes in routes of some bus services. 2) The bus company temporarily provides additional bus services to cater for special events. 3) We only retain the data of bus routes that we are able to infer without any errors.
\begin{table}[htb]
	\centering
	\caption{Statistics of Transportation Network}
	\label{tbl:statistics}
	\begin{tabular}{|p{2.1cm}|r|r|r|}
		\hline
			& Dec 8th & Dec 15th & Dec 22nd \\
		\hline
		No. of records & 3,137,469 & 3,176,830 & 3,162,985 \\
		\hline
		No. of inferred bus routes & 288 & 285 & 287 \\
		\hline
		No. of nodes & 4497 & 4492 & 4514 \\
		\hline
		No. of edges & 6176 & 6151 & 6195 \\
		\hline
	\end{tabular}
\end{table}

\subsection{Sum-of-Squares Error Convergence}

We would first like to verify that the Stochastic Gradient Descent (SGD) algorithm converges for our proposed model in Section \ref{sec:models}. The SGD algorithm is supposed to reduce the error between the expected time given by the models and observed time as reflected in the records $r \in R$. The error for all records $r \in R$ is given by the Sum-of-Squares Error (SSE) as shown in Equation \ref{eqn:sum_of_squares}. A lower SSE value suggests a better fit of the model to the observed data.
\begin{align}
	\label{eqn:sum_of_squares}
	\text{Sum-of-Squares Error} = \sum_{r \in R} (t_r - \hat{t}_r)^2 
\end{align}

\begin{figure}[htb]
	\centering
	\includegraphics[width=3.0in]{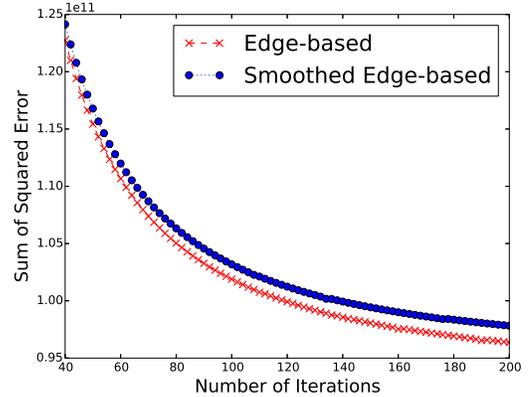}
	\caption{Stochastic Gradient Descent Convergence rate for Sum-of-Squares Error}
	\label{fig:convergence}
\end{figure}

Figure \ref{fig:convergence} shows the convergence of SSE with respect to the number of iterations we ran for the SGD algorithm. As observed in Figure \ref{fig:convergence}, the SGD algorithm decreases the SSE over multiple iterations with the Edge-based model having a lower SSE than Smoothed Edge-based model. This is due to the additional smoothing constraints we have imposed on the Smoothed Edge-based model. The change in SSE decreases as number of iterations increase, which suggests that the SGD algorithm converges for our two proposed models, resulting in marginal improvement in the SSE with increasing number of iterations.

\subsection{K-Fold Cross Validation}

We evaluate the two models proposed in Section \ref{sec:models} on how well they generalize to data that is not seen during the training (inference and estimation) phase. We perform a $K$-fold cross validation evaluation on each day of data. The procedure for $K$-fold cross validation first uniformly and randomly divide the data set into $K$ number of smaller data sets called folds. We choose $K-1$ of these folds to represent the observed data $R_{train}$ for training, while the remaining fold is used as the unseen data $R_{test}$ for testing the goodness-of-fit of the estimated parameters obtained after training. The goodness-of-fit we used in this part of the experiments is the Root-Mean-Squared-Error (RMSE).
\begin{align*}
	\text{RMSE of data} = \sqrt{\frac{\sum_{r \in R_{data}} (t_r - \hat{t}_r)^2}{|R_{data}|}}
\end{align*}

\begin{figure}[htb]
	\centering
	\subfloat[Dec 8th 2011]{
		\includegraphics[width=1.7in]{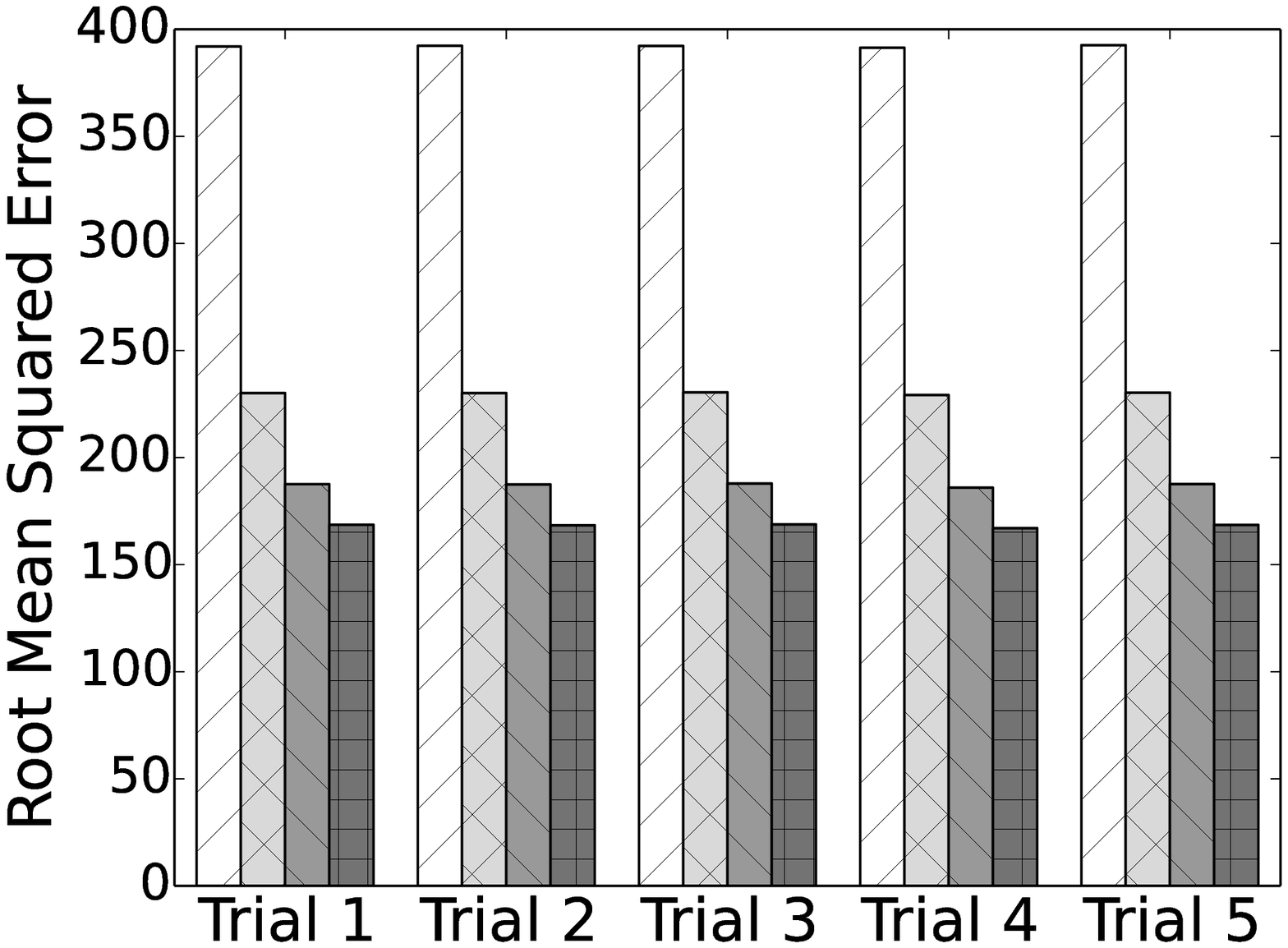}
		\label{fig:20111208_train}
	}
	\subfloat[Dec 15th 2011]{
		\includegraphics[width=1.7in]{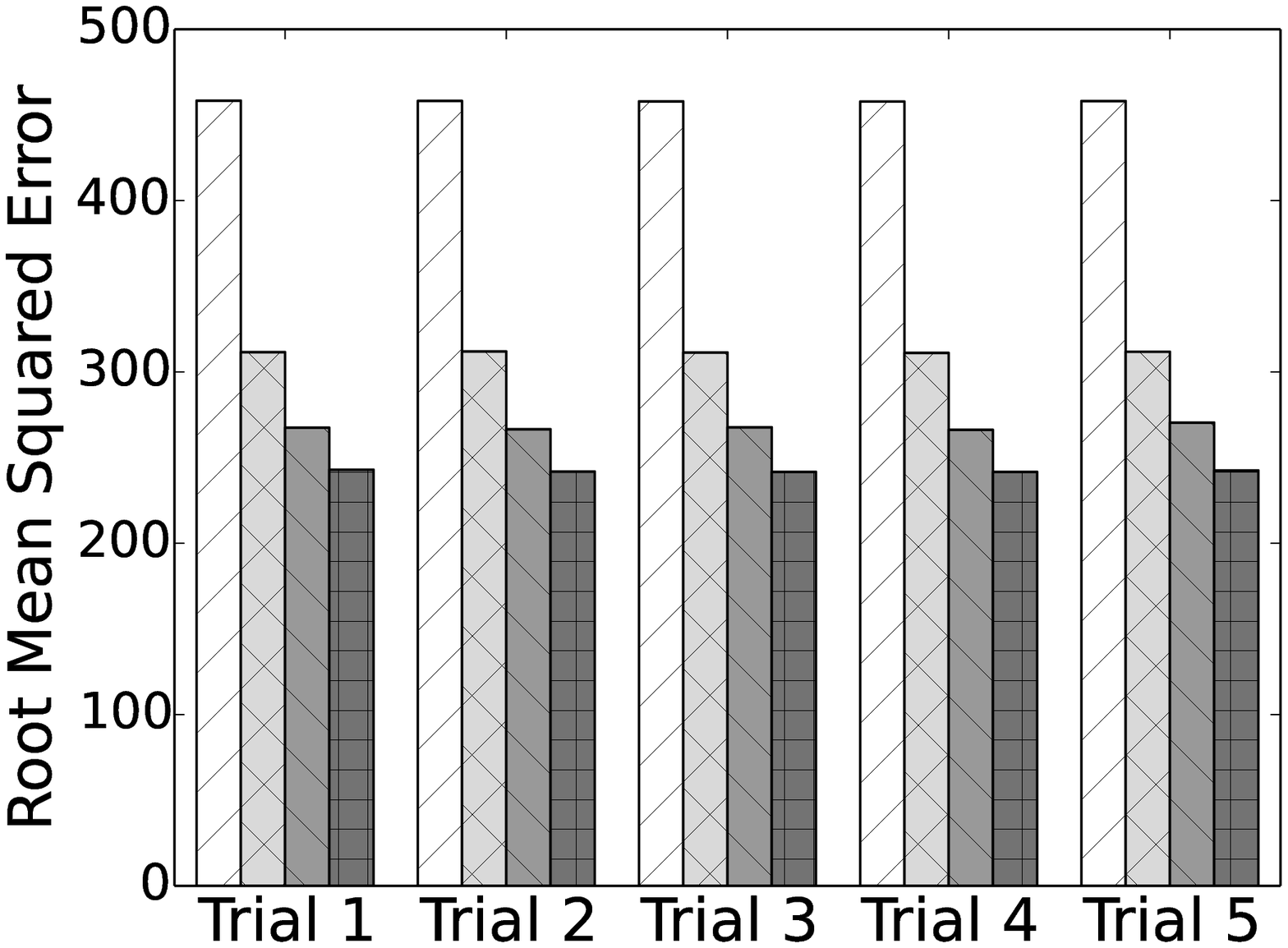}
		\label{fig:20111215_train}
	}\\
	\subfloat[Dec 22nd 2011]{
		\includegraphics[width=1.7in]{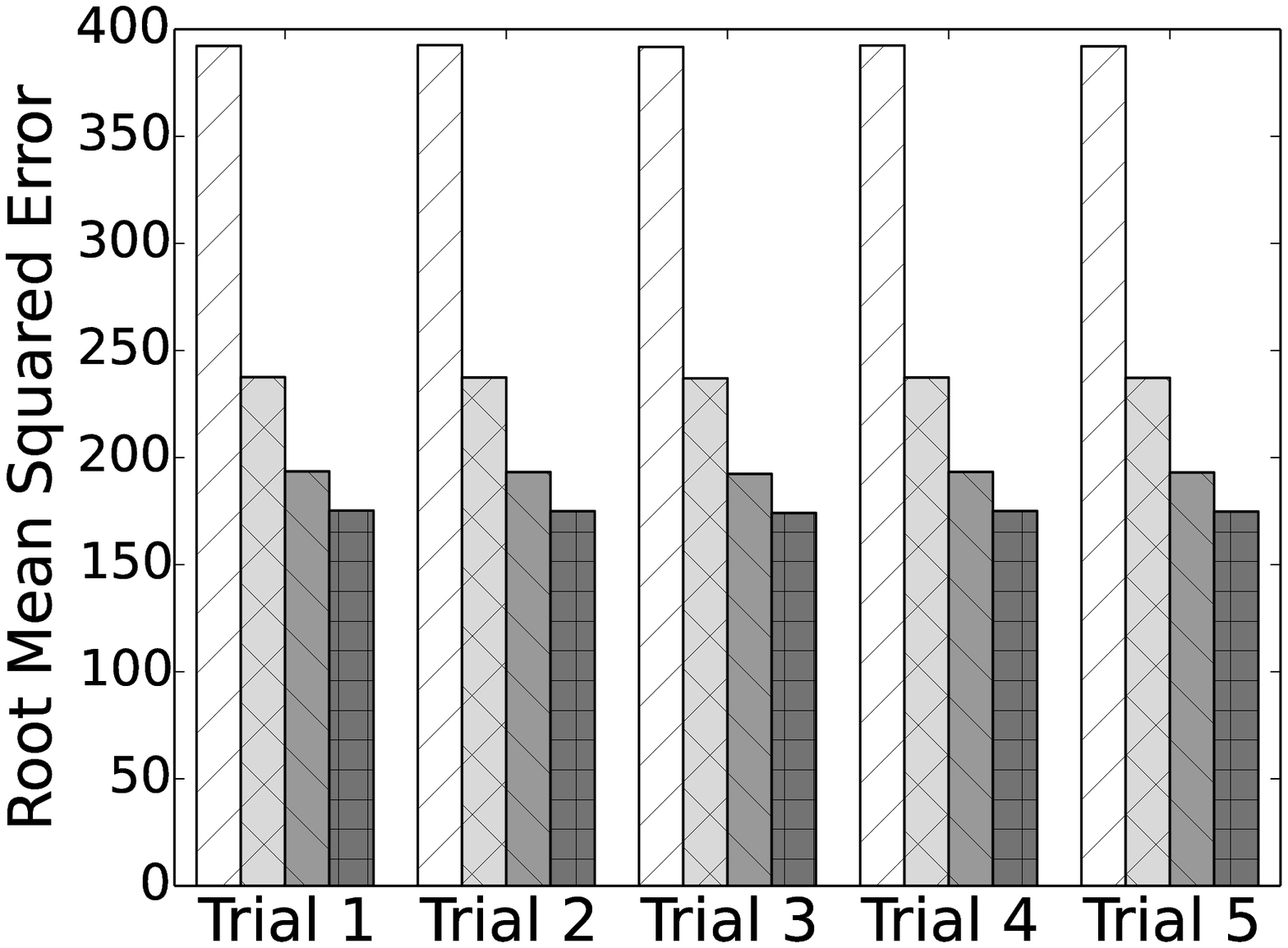}
		\label{fig:20111222_train}
	}
	\subfloat[Legend]{
		\includegraphics[width=1.7in]{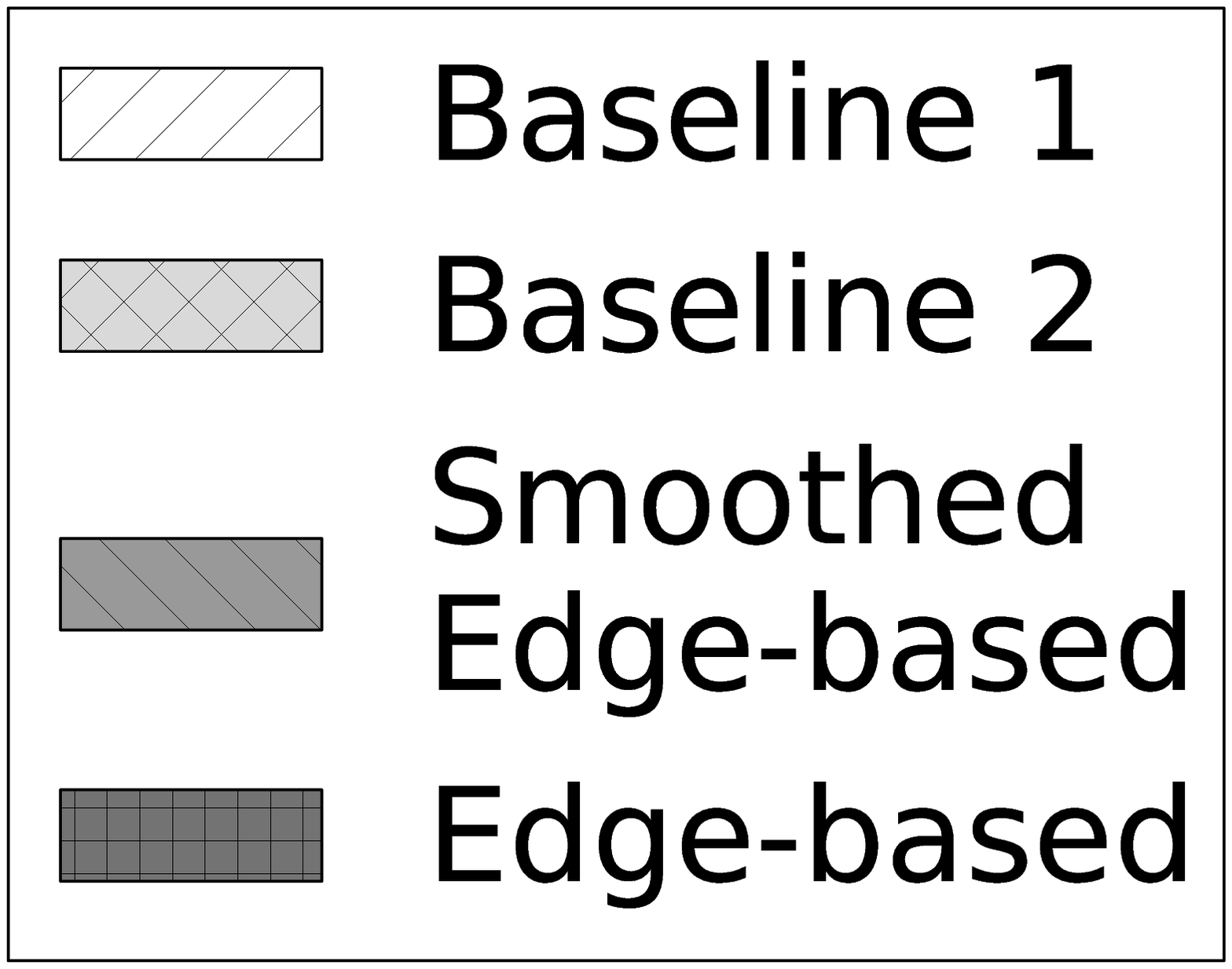}
		\label{fig:legend_train}
	}
	\caption{Training Set: 5 Fold Cross Validation}
	\label{fig:5_fold_train}
\end{figure}

\begin{figure}[htb]
	\centering
	\subfloat[Dec 8th 2011]{
		\includegraphics[width=1.7in]{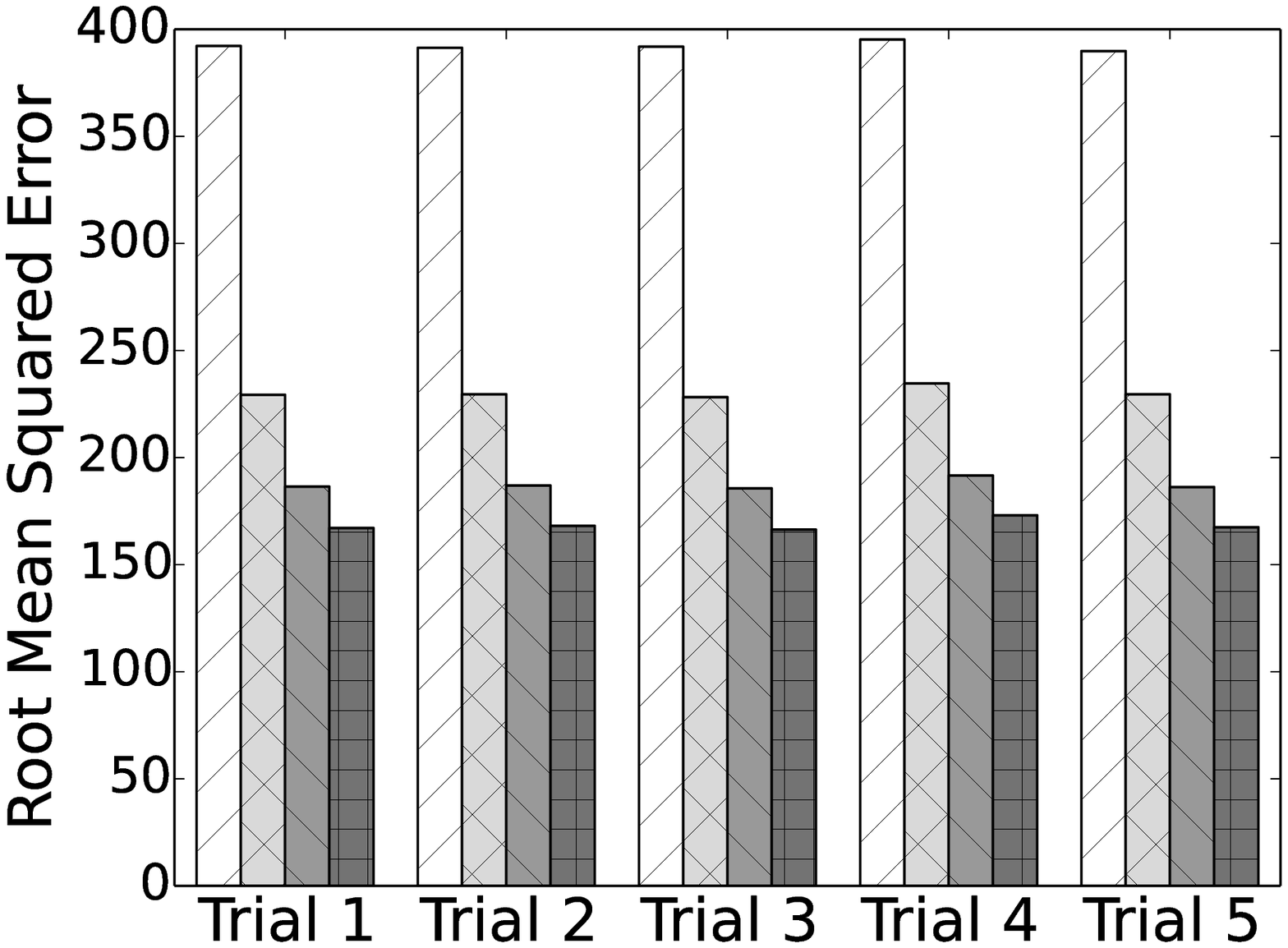}
		\label{fig:20111208_test}
	}
	\subfloat[Dec 15th 2011]{
		\includegraphics[width=1.7in]{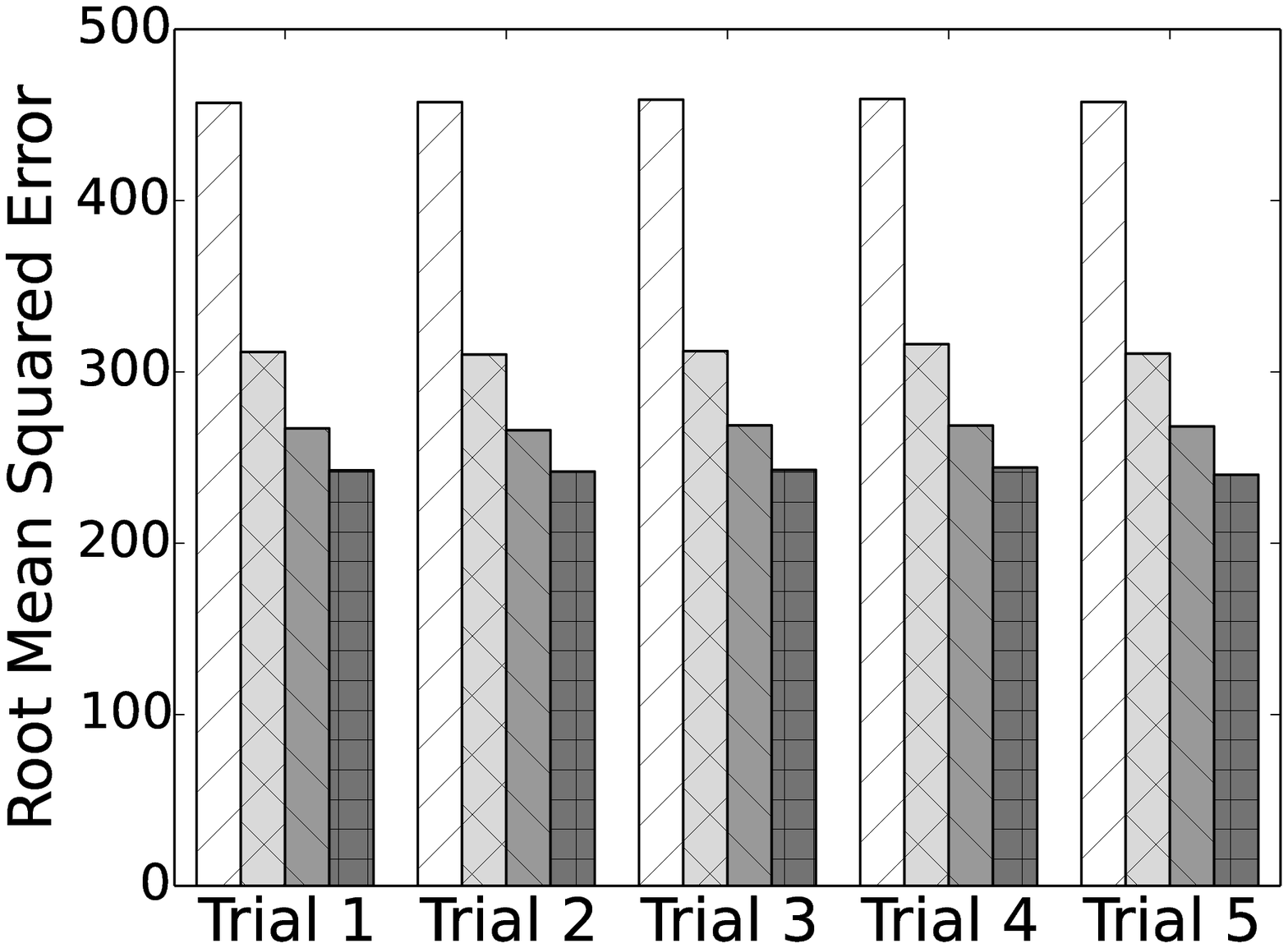}
		\label{fig:20111215_test}
	}\\
	\subfloat[Dec 22nd 2011]{
		\includegraphics[width=1.7in]{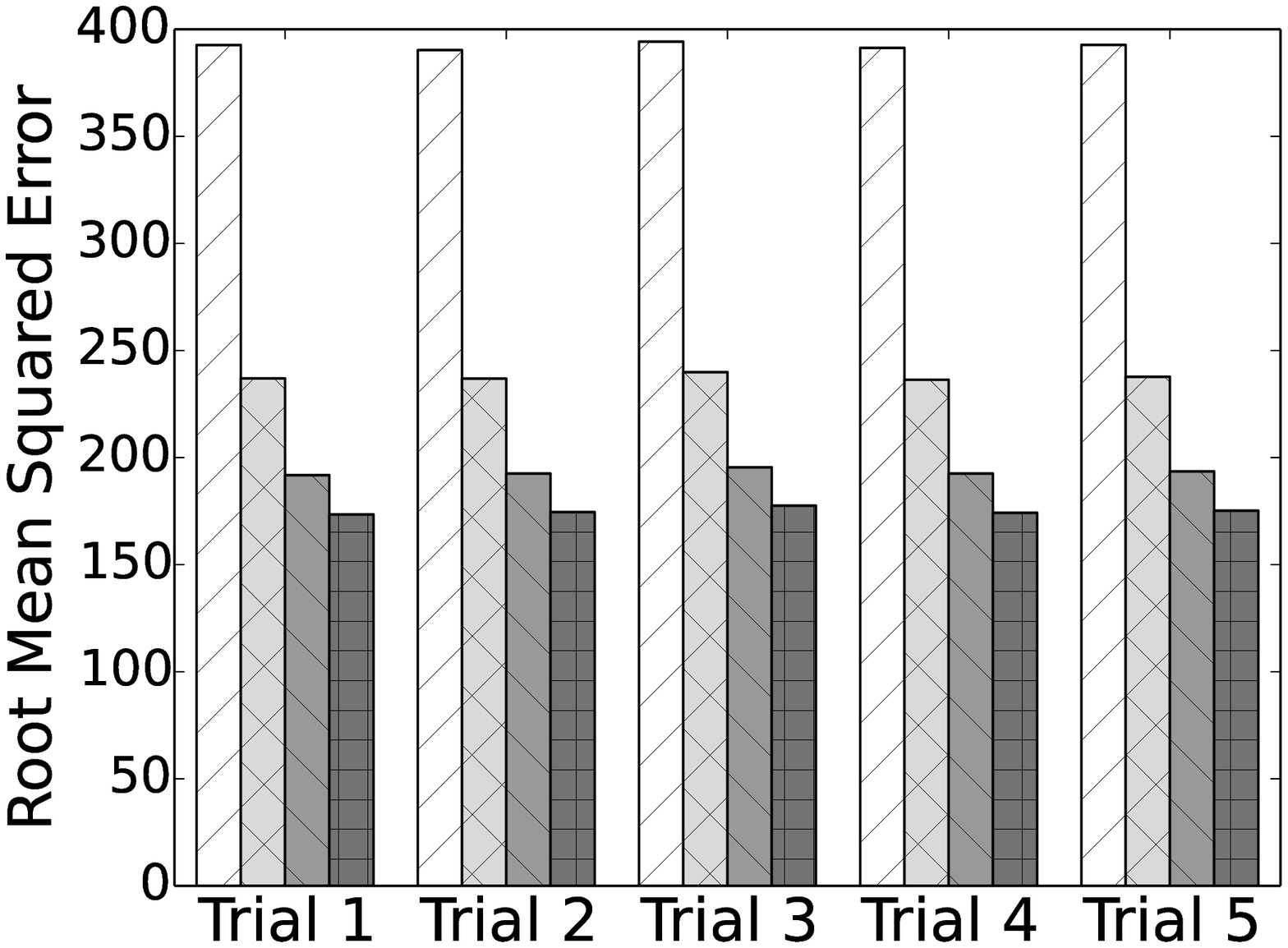}
		\label{fig:20111222_test}
	}
	\subfloat[Legend]{
		\includegraphics[width=1.7in]{legend}
		\label{fig:legend_test}
	}
	\caption{Testing Set: 5 Fold Cross Validation}
	\label{fig:5_fold_test}
\end{figure}

The experiment is repeated for $K$ number of trials, by cycling the test set through each of the $K$ folds. We performed the K-fold cross validation independently on three days of data and obtain the results as shown in Figures \ref{fig:5_fold_train} and \ref{fig:5_fold_test}. 

Figures \ref{fig:20111208_train} to \ref{fig:20111222_train} show the RMSE of the data set $R_{train}$ used for estimating the parameters of the models in Section \ref{sec:models}. Using the estimated parameters, we proceed to obtain the expected time taken for the records $r \in R_{test}$, and obtain the respective RMSE of the test data, shown in Figures \ref{fig:20111208_test} to \ref{fig:20111222_test}.

In all of these Figures, with reference to the legend shown in Figures \ref{fig:legend_train} and \ref{fig:legend_test}, we illustrate the RMSE given by various models for each trial. The results consistently show that the Edge-based model has the best performance, follow by Smoothed Edge-based, then Baseline 2 and finally Baseline 1. This suggests that two of our proposed models outperform the baselines and the speeds we have inferred for the segments are accurate in estimating the expected times of the journeys. Given that Edge-based model outperforms Smoothed Edge-based model, we will use the results of the Edge-based model for the rest of our analysis and ignore the smoothing constraints of the Smoothed Edge-based model.

We also observed that the RMSE of all four models in Figures \ref{fig:20111215_train} and \ref{fig:20111215_test} for Dec 15th 2011 is higher than the RMSE shown the Figures representing Dec 8th 2011 (Figures \ref{fig:20111208_train}, \ref{fig:20111208_test}) and Dec 22nd 2011 (Figures \ref{fig:20111222_train} and \ref{fig:20111222_test}). This is an indication that a significant anomaly is present on the day of Dec 15th 2011, which causes the RMSE goodness-of-fit for our proposed model to differ from other normal days. In the next section, we will give a case study of identifying the anomalies using the algorithm we proposed in Section \ref{sec:localization}.

\subsection{Evaluation of Anomalies using Twitter}

The main objective which motivates our research is to detect and localize the network anomalies in distributed systems. Given that in comparison to other models, the Edge-based model gives a more accurate estimation of how long a journey should take, we use the estimation of Edge-based model as inputs to the algorithm described in Section \ref{sec:localization} to detect and localize the network anomalies.

We mentioned in Section \ref{sec:intro} that we focus on non-critical anomalies which are elusive and difficult to detect in real usage of distributed systems. As a result of this elusive property of non-critical anomalies, our detection and localization algorithm does not have sufficient labeled data to evaluate its accuracy in a quantitative manner. In fact, there would be no research problem to address if such anomalies are easily observed for us to perform our experiments.

Fortunately, since we use the Public Transportation System (PTS) of an urbanized city, we are able to compare the detected anomalies with the tweets of the residents who commute in the city. We use the Twitter data set that was formerly used to analyze political sentiments in Hoang and Lim \cite{Hoang2012}, to perform a qualitative comparison of our detected anomalies with the tweets that comment on the traffic conditions. We use the Twitter data set $\mathcal{T}$, and the PTS records $R$ between the periods of November 1st 2011 and January 31st 2012 for comparison.

We described in Section \ref{sec:localization} that the detection algorithm first obtains the set of records $R_{\alpha > \delta}$, where each record $r \in R_{\alpha > \delta}$ has a larger observed travel time than the expected travel time, i.e. $\alpha_r > \delta$. We set $\delta$, the cut-off value of determining whether a record deviates significantly as the top 1\% of the ratio values $\alpha_r$, for all $r \in R$. The records $R_{\alpha > \delta}$ are then sorted in descending order of $|R_r|$, so that records $r \in R_{\alpha > \delta}$ with the most important path $p_r$ are ranked first. 

\subsubsection{Evaluation of $|R_r|$ vs $\alpha_r$}

We evaluate the use of $|R_r|$ vs $\alpha_r$ as a ranking metric by comparing with the number of tweets in $\mathcal{T}$ that mention the keywords (ignoring case) $\mathcal{K}$, 
\begin{center}
	$\mathcal{K}$ := \{ ``traffic'' $\land$ (``jam'' $\lor$ ``jams'') \}.
\end{center}
Table \ref{tbl:twitter} provides the statistics of the Twitter data $\mathcal{T}$ and the tweets $\mathcal{T}_{\mathcal{K}}$ with the keywords $K$ for each month. Table \ref{tbl:twitter} also show the number of tweets $\mathcal{T}_{\mathcal{M}}$ containing keywords $\mathcal{M}$, which we will elaborate in later part of this section.
\begin{table}[htb]
	\centering
	\caption{Statistics of Twitter Data}
	\label{tbl:twitter}
	\begin{tabular}{|l|r|r|r|}
		\hline
			Period & $| \mathcal{T} |$ & $| \mathcal{T}_{\mathcal{K}} |$ & $| \mathcal{T}_{\mathcal{M}} |$ \\
		\hline
			Nov 2011 & 17,421,755 & 1,330 & 36 \\
		\hline
			Dec 2011 & 19,216,767 & 1,829 & 8,247 \\
		\hline
			Jan 2012 &  19,565,979 & 1,891 & 847 \\
		\hline
	\end{tabular}	
\end{table}

\begin{figure}[htb]
	\centering
	\subfloat[Histograms of $\mathcal{T}_{\mathcal{K}}$ that mention ``traffic jam(s)'']{
		\includegraphics[width=3.5in]{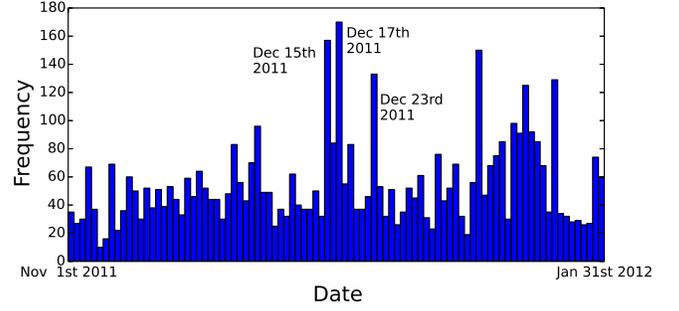}
		\label{fig:twitter_traffic_jam}
	}\\
	\subfloat[$|R_r|$: Mean values has 0.64 Pearson correlation with frequencies]{
		\includegraphics[width=3.5in]{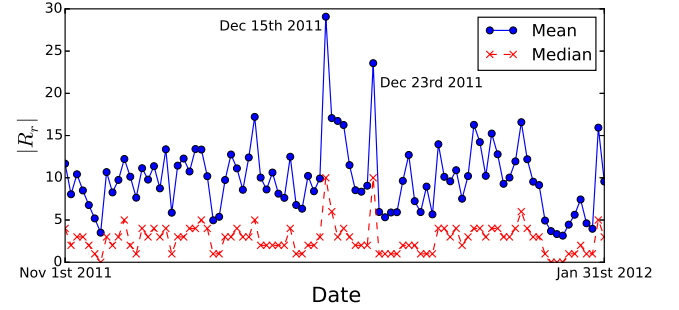}
		\label{fig:Rr_mean_median}
	}\\
	\subfloat[$\alpha_r$: Mean values has 0.20 Pearson correlation with frequencies]{
		\includegraphics[width=3.5in]{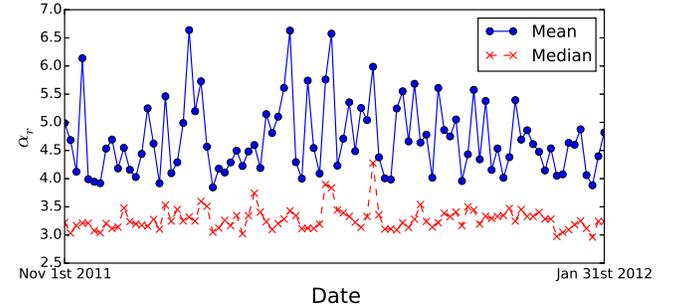}
		\label{fig:alpha_r_mean_median}
	}
	\caption{Comparision of Anomaly Ranking Metric between $|R_r|$ and $\alpha_r$ for the period between Nov 1st 2011 and Jan 1st 2012}
	\label{fig:Rr_vs_alpha_r}
\end{figure}

Figure \ref{fig:twitter_traffic_jam} shows the histogram of tweets $\mathcal{T}_{\mathcal{K}}$ containing keywords $\mathcal{K}$. The width of the bins in Figure \ref{fig:twitter_traffic_jam} is chosen to represent the duration of one day, so that the frequency (y-axis) shown in the histogram represents the number of tweets that contain keywords $\mathcal{K}$ for the specific day (x-axis).

We used the Edge-based model of Section \ref{sec:models} and the anomalies localization algorithm of Section \ref{sec:localization} on the records of each day between the periods of November 1st 2011 to January 1st 2012. For each day, we derive $\alpha_r$ of every record $r$ in that day using Equation \ref{eqn:alpha_ratio}, then we obtain the set of records $R_{\alpha > \delta}$, and derive the set of $R_r, \forall r \in R_{\alpha > \delta}$. Next, we obtain the mean and median values of $|R_r|, \alpha_r, \forall r \in R_{\alpha > \delta}$ of each day and derive the plots shown in Figures \ref{fig:Rr_mean_median} and \ref{fig:alpha_r_mean_median}. 

The histogram in Figure \ref{fig:twitter_traffic_jam} appear to correlate better with the plots in Figure \ref{fig:Rr_mean_median} compared to the plots in Figure \ref{fig:alpha_r_mean_median}, especially for the two dates marked in Figure \ref{fig:Rr_mean_median}. We calculated the Pearson correlation of the frequencies in Figure \ref{fig:twitter_traffic_jam} with the mean values in Figure \ref{fig:Rr_mean_median} and obtain the value of 0.64. On the other hand, the Pearson correlation between the frequencies of Figure \ref{fig:twitter_traffic_jam} and mean values of Figure \ref{fig:alpha_r_mean_median} is only 0.20. This confirms that our proposed metric of $|R_r|$ is better than $\alpha_r$ for ranking the anomalies in the set of records $R_{\alpha > \delta}$.

\subsubsection{Case studies of the location for the detected\\anomalies}

In this final section of our evaluation on the detected anomalies, we describe the details of evaluating the location and time of our detected anomalies through qualitative comparison with the contents of the tweets. From Figure \ref{fig:twitter_traffic_jam}, we have marked December 15th and 17th of 2011 which shows significant spike in the number of tweets mentioning ``traffic jam(s)''. Based on prior knowledge, we are aware of the existence of an external event that causes the congestion on these two days. The external event is the breakdown of the railway system that caused temporary disruption to the railway services.

The city is supported by two main transportation systems, the railway system (MRT\footnote{MRT is the acronym of the railway system and it stands for Mass Rapid Transit.}) and the public bus system. When the railway system breaks down, the remaining transportation options in the city has to bear the load of transporting the passengers. These other transportation options which include taxis and privately owned motorized vehicles share the well-connected road network with the public bus system. The sharing of physical road space causes delay in the speeds of bus during the breakdown of the railway system. 

\begin{figure}[htb]
	\centering
	\subfloat[Frequency for each Day]{
		\includegraphics[width=3.5in]{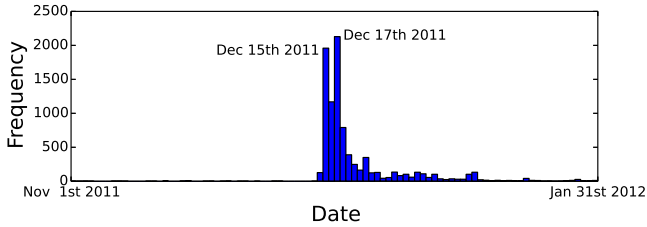}
		\label{fig:twitter_mrt_breakdown}
	}\\
	\subfloat[Log(Frequency) for each Day]{
		\includegraphics[width=3.5in]{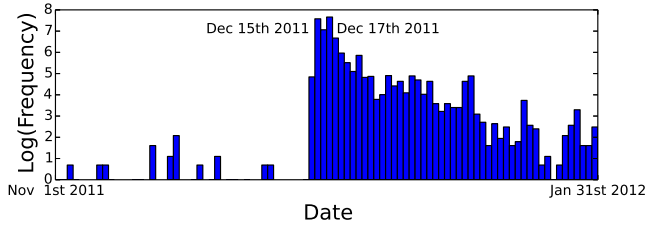}
		\label{fig:twitter_mrt_breakdown_log}
	}
	\caption{Histograms of tweets that mention ``mrt breakdown'' or ``mrt break(s) down''}
	\label{fig:twitter_histograms}
\end{figure}

In order to determine when the railway system breaks down, we search the set of tweets $\mathcal{T}$ between the periods of November 1st 2011 to January 31st 2012 for the keywords (ignoring case) $\mathcal{M}$,
\begin{center}
	$\mathcal{M}$ := \{ mrt $\land$ (break down $\lor$ breaks down $\lor$ breakdown) \}
\end{center}
and obtain the subset of tweets $\mathcal{T}_\mathcal{M}$. Table \ref{tbl:twitter} shows statistics of $\mathcal{T}_\mathcal{M}$ for each month. We use Figures \ref{fig:twitter_mrt_breakdown} and \ref{fig:twitter_mrt_breakdown_log} to show the number of tweets generated for each day. Figure \ref{fig:twitter_mrt_breakdown} shows that there are two days, May 15th 2011 and May 17th 2011 with many tweets containing the keywords $\mathcal{K}$. These two days also correlate with the frequencies as seen in Figure \ref{fig:twitter_traffic_jam}.

\begin{table*}[htb]
	\centering
	\caption{Examples of tweets from $\mathcal{T}_{\mathcal{M}}$, commenting on the traffic after railway system (MRT) breaks down}
	\label{tbl:tweets}
	\begin{tabular}{|r|p{4cm}|p{9.5cm}|p{2.1cm}|}
		\hline
		\# & URL & Content & Time GMT+8 \\
		\hline
		1 & \url{http://twitter.com/tickingbombs/status/131164834623524864} & Train breakdown at \textbf{DS}. Whats the fare hike for again? & 2011-11-01 7:25:08  \\
		\hline
		2 & \url{http://twitter.com/mrbrown/status/146756643298877440} & ... breakdown pic \#smrt RT @hai\_ren: The crowd at \textbf{PL}. \#SMRTruinslives \url{http://t.co/RGQwSp3s} & 2011-12-14 8:01:25 \\
		\hline
		3 & \url{http://twitter.com/mac_beno_l/status/147304462514520064} & Major MRT break down, causing MAJOR traffic jam along \textbf{OR}. Feeling very sick :( & 2011-12-15 20:18:15 \\
		\hline
		4 & \url{http://twitter.com/true_joygiver/status/147319951831736320} & I'm a victim of the serious MRT breakdown! Stuck in \textbf{SS} 4 hours \& forced to miss a dinner gathering! =( & 2011-12-15 21:19:48 \\
		\hline
		5 & \url{http://twitter.com/ONGLSD/status/147323153243324418} & The train breakdown isn't just affecting SMRT's traffic, it's causing a major killer jam from \textbf{SR} all the way to \textbf{OR}. AVOID!! & 2011-12-15 21:32:31 \\
		\hline
		6 & \url{http://twitter.com/afiqahmauwan/status/147367254059782144}	& That MRT breakdown took me 2.5 hours to reach home from \textbf{DG} to \textbf{YI}. & 2011-12-16 00:27:45 \\
		\hline
		7 & \url{http://twitter.com/aidah/status/147824290526535681} & MRT breakdown this morning between \textbf{MB} and \textbf{NS} \#fb \#north-south line. Frustrating. & 2011-12-17 6:43:51 \\
		\hline
	\end{tabular}
\end{table*}

\begin{table*}[htb]
	\centering
	\caption{Summary of Anomalies}
	\label{tbl:anomalies}
	\begin{tabular}{|r|r|r|p{2.3cm}|p{2.3cm}|p{1.2cm}|p{1.6cm}|p{1.6cm}|p{1.4cm}|p{1.4cm}|}
		\hline
		\# & $|R_r|$ & $\alpha_r$ & Origin & Destination & Distance (meters) & Time Boarded & Time Alighted & Observed (mins) & Expected (mins) \\
		\hline
		1 & 46 & 3.89 & N.A. & DS & 2,900 & Nov 1 2011 08:04:37 & Nov 1 2011 08:23:34 & 18.95 & 9.11 \\
		\hline
		2 & 69 & 3.16 & PL & N.A & 600 & Dec 14 2011 08:43:35 & Dec 14 2011 08:49:57 &6.37 & 3.11\\
		\hline
		3 & 483 & 5.79 & SS & OR & 1,000 & Dec 15 2011 20:54:01 & Dec 15 2011 21:15:04 & 21.05 & 10.22 \\
		\hline
		4 & 181 & 25.27 & SR & SS & 1,700 & Dec 15 2011 20:14:37 & Dec 15 2011 21:32:02 & 77.42 & 15.79 \\
		\hline
		5 & 161 & 14.12 & SR & OR & 2,700 & Dec 15 2011 07:57:14 & Dec 15 2011 21:06:36 & 69.37 & 26.01 \\
		\hline
		6 & 0 & 4.16 & DG & YI & 31,900 & Dec 15 2011 20:48:23 & Dec 15 2011 23:17:43 & 149.33 & 105.38 \\
		\hline
		7 & 149 & 12.98 & NS & N.A. & 200 & Dec 17 2011 11:54:52 & Dec 17 2011 12:07:45 & 12.88 & 4.51 \\
		\hline
	\end{tabular}
\end{table*}

Table \ref{tbl:tweets} shows examples of tweets that comment on the traffic conditions whenever the railway system breaks down, with Tweets \#3 and \#5 of Table \ref{tbl:tweets} showing passengers' feedback that the breakdown of railway system also affects the bus travel times. At the time of this writing, all the URLs as shown, link to the publicly available tweets on Twitter website. In each of these tweets, the railway stations affected by the breakdown that are mentioned by the users are highlighted in \textbf{bolded font}\footnote{Due to requests from our data providers, we have to anonymize the location names with their initials.}.

Although we have the complete record of passengers for the three months listed, we do not have the complete set of tweets during this period. We also do not assume that Twitter is able to capture all the passengers' feedback about the usage of the public bus system during this period of time. It is therefore not possible to obtain a quantitative correlation between the ranking of anomalies detected by our algorithm and the number of tweets mentioning the location of the congested areas. But from the content of tweets shown in Table \ref{tbl:tweets}, we can find specific records with ratio $\alpha_r$ that belong to the top 1\% and match the description of the tweets.

We use Table \ref{tbl:anomalies} to show examples of records that correspond in terms of location, date and time to the traffic congestions as mentioned by Twitter users in Table \ref{tbl:tweets}. Each record \# in Table \ref{tbl:anomalies} corresponds to the tweet \# in Table \ref{tbl:tweets}. One may notice that either the origin or destination for each record in Table \ref{tbl:anomalies} is related to the bolded location of the tweets in Table \ref{tbl:tweets}, while N.A. indicates that it is not related. The times of the record are also very close to the tweets and we could assume that passengers\footnote{Disclaimer: The passengers in the records of Table \ref{tbl:anomalies} are not the same people as the Twitter users of Table \ref{tbl:tweets}.} tweet about their frustration during or after the journey. The only exception is \#7 that has a huge difference in times because Dec 17 is a Saturday (non-working day) and passengers only commute later in the day. The records in Table \ref{tbl:anomalies} show that the observed time taken to reach the destination is much longer than the expect time. While these few examples do not necessarily cover all cases, it certainly shows that our detected anomalies could match with the complaints in Twitter or any other social media.

\section{Conclusion}
\label{sec:conclusion}

We began our research with the purpose of finding non-critical anomalies in the networks of distributed systems. In most distributed systems, the data that can be obtained without the use of sophisticated instruments or internal sensors is often non-informative about the internal workings of the networks. That causes difficulties in detecting anomalies and further difficulties in finding the location of anomalies within the distributed system. To overcome these difficulties, we proposed the Edge-based network transmission model to infer the flow speeds of the edges within the networks of distributed systems. With the model, we are able to derive the expected time necessary for entity to complete its flow. Using the records of entities flow with observed time that is significantly longer than expected, we apply our proposed localization algorithm to measure the relationship of each record to all other records with large deviations. The number of related records allows us to determine how important each record is to the traffic conditions of the network. By finding records that are highly related to other records and with the shortest travel path in the network, we are able to determine the location of the anomalies within the distributed system.

One major assumption that we had made in this work is that the knowledge of the exact path taken by the entity flow is known or can be easily inferred. While this is true for transportation systems, it may not be true for other kinds of distributed systems. One way to overcome this is to have an intermediate step to infer the path using a Markovian model. 

We make some final remarks about other possible improvements. The current models have not considered the notion of peak and off-peak usage of the traffic patterns in networks of distributed systems. During peak usage, the load is generally higher and could result in longer observed travel times. This issue can be easily addressed by using a mixture of Gaussian distributions to model the edge speeds. Another possible improvement is the algorithm for counting the ``contains'' and ``within'' relationships between records. Using concepts of transitivity, e.g. if record $a$ contains record $b$, and record $b$ contains record $c$, then record $a$ contains record $c$, one could save on the computation costs significantly.

\section*{Acknowledgment}
We would like to thank the Land Transport Authority (LTA) of Singapore for sharing with us the EZLink dataset.

\bibliographystyle{IEEEtran}
\bibliography{references}

\end{document}